\documentstyle[preprint,aps]{revtex}

\newcommand{\disregard}[1]{}

\newcommand{\be}{\begin{equation}}
\newcommand{\ee}{\end{equation}}
\newcommand{\rmT}{T}
\tighten
\begin{document}

\draft

\title{Shell  Structure of the Superheavy Elements}

\author  {
S. \'Cwiok$^{1,2}$,
J. Dobaczewski$^{1,3}$,
P.-H. Heenen$^{4}$,
P. Magierski$^{2}$,
 W. Nazarewicz$^{3,5,6}$}

\address {
$^1$Joint Institute for Heavy Ion Research,
   Oak Ridge National Laboratory                                \\[-1mm]
   P.O. Box 2008, Oak Ridge,   TN 37831, U.S.A.                 \\
$^2$Institute of Physics,
Warsaw University of Technology                  \\[-1mm]
ul. Koszykowa 75, PL-00662, Warsaw, Poland\\
$^3$Institute of Theoretical Physics, Warsaw University         \\[-1mm]
   Ho\.za 69,  PL-00681, Warsaw, Poland                         \\
$^4$Service de Physique Nucl\'{e}aire Th\'{e}orique \\[-1mm]
    U.L.B - C.P. 229, B-1050 Brussels, Belgium \\
$^5$Department of Physics, University of Tennessee,
                                     Knoxville, TN 37996, U.S.A.\\
$^6$Physics Division, Oak Ridge National Laboratory             \\[-1mm]
   P.O. Box 2008, Oak Ridge,   TN 37831, U.S.A.                 \\
}
\maketitle

\begin{abstract}
Ground state properties of  the superheavy elements (SHE)  with
108$\leq$$Z$$\leq$128 and 150$\leq$$N$$\leq$192 are investigated  using
both
the Skyrme-Hartree-Fock method with a density-independent contact pairing
interaction and
the macroscopic-microscopic approach  with  an  average  Woods-Saxon
potential and a monopole pairing interaction. Detailed analysis
of binding energies, separation energies, shell effects,  single proton
and neutron states, equilibrium deformations,
 $Q_\alpha$-values, and other observables is given.
\end{abstract}


\narrowtext

\section{Introduction}

Recent  years have
 witnessed great strides in the production
of the heaviest nuclei.
Notably, three new elements, $Z$=110, 111, and 112,
were synthesized at GSI
\cite{[Hof95],[Hof96]},
 Berkeley \cite{[Ghi95]}, and  Dubna \cite{[Oga95]}.
The systematics of reaction data
show that cross-sections are monotonically
decreasing, reaching a level of 1\,pb for production of element
112. The
half-lives of the known isotopes of elements with $Z$$>$105
are predominantly determined by the $\alpha$-decay, and decrease from
0.9\,s for $^{263}$106 to 0.2\,ms
 for $^{277}{112}_{165}$.
These isotopes  decay predominantly by
 groups of
$\alpha$ particles ($\alpha$-chains).
The corresponding  $\alpha$-spectra
carry   important
nuclear structure information.

 All the heaviest  elements  found recently
 are believed to be well deformed.
Indeed, the measured $\alpha$-decay energies,  along with
complementary syntheses of new neutron-rich isotopes of elements $Z$=106 and
$Z$=108 by the Dubna-Livermore collaboration, have furnished confirmation of
 the special stability of the deformed
shell at $N$=162
\cite{[Cwi83],[Cwi85],[Bon86a],[Sob87],[Sob89],[Pat89],%
[Cwi92],[Laz94],[Cwi94a],[Mol94a],[Smo95]}.
 Still heavier and more neutron-rich
elements  are expected to be spherical and even more strongly
stabilized by shell effects;
they form the long predicted region of the superheavy elements (SHE).
A long-standing goal in
themselves, SHE are  also a critical test of nuclear models,
 of relativistic effects, and of quantum chemistry.

Experimentally, the excursion towards the center of the SHE will
not be easy. Indeed,
for the most neutron-rich
isotopes synthesized, production cross sections are of the order
of pb which is on the
limit of today's experimental sensitivity.
Two ways of synthesizing new elements have been employed.
Experiments at GSI
 utilized  cold fusion reactions
$^{208}$Pb(HI, xn) which
 benefit from the enhanced stability of the target nuclei.
 On the other hand, the Dubna-Livermore
experiments utilized  more asymmetric hot fusion
reactions
$^{244}$Pu(HI,xn). As far as  future measurements are concerned,
there have been intensive discussions on how to reach
still-heavier new
nuclei \cite{[Oga95],[Mue88],[Kum90],[Sch91a],[Hof95a]}.
 Though the experiments will be difficult, and will
probably require advances in separation and detection technology, 
one advantage is that many
of the nuclei sought are predicted to have longer lifetimes
 than those recently synthesized.
Clearly, there is now a wealth of possibilities opening up involving cold and
hot fusion, light and heavy projectiles,
and  stable and radioactive beams.

Among various theoretical approaches to the heaviest elements,
particularly successful have been calculations
based on the macroscopic-microscopic
method (Nilsson-Strutinsky approach). In particular,
detailed studies have been carried out
with the folded-Yukawa (FY) deformed single-particle potential
\cite
{[Mol81],[Mol81a],[Ben84],[Mol86],[Mol88],[Mol88a],%
[Mol92a],[Mol95]}
and
with the Woods-Saxon (WS) deformed single-particle potential
\cite
{[Cwi83],[Cwi85],[Bon86a],[Sob87],[Sob89],[Pat89],[Cwi92],[Cwi94a],[Smo95],%
[Pat89a],[Pat91]}.
In most cases, the FY and WS results are consistent; they both
reproduce well-known alpha-decay  half-lives of  heavy
elements and
 emphasize the stabilizing role of
deformation for  nuclei with N$\approx$162.
Since the liquid-drop barriers in SHE  disappear,
the shell effects dramatically influence the spontaneous fission
and alpha-decay half-lives; the delay in the spontaneous fission
half-lives, ${\rmT}_{\rm sf}$,
 due to the shell effects is  about 15 orders of magnitude
for Z$\gtrsim$106 \cite{[Mue88]}. In this context, it is important to recall that
in both FY and WS models
the center of the shell  stability
is predicted around the  alleged spherical doubly-magic nucleus
$^{298}$114$_{184}$.

The 
 aim of the present study is to  understand  the
shell structure  properties of SHE based on the self-consistent theory.
Although the
physics  of the heaviest and superheavy nuclei has
been  addressed by
numerous  calculations (for a review, see Refs.
\cite{[Mol94a],[Kum89]}), a large-scale self-consistent description
of SHE
involving non-spherical  degrees of freedom is still lacking.
It is our intention to fill this gap. The
macroscopic-microscopic approach, which has mostly been used up to now,
is only
an approximation to the Hartree-Fock (HF) theory and suffers from
the lack of self-consistent coupling between
the  macroscopic and microscopic
contributions to the total energy. Moreover, it is not based on a
microscopic  many-body
 Hamiltonian and its
success (or failure) in reproducing experimental data cannot be easily
traced back to an effective interaction (or to the corresponding
density functional).
In this  work we
analyze the properties of the SHE by means of the
self-consistent Skyrme-Hartree-Fock approach
with a zero-range  pairing interaction.
For comparison, a set of calculations has also been carried out
with the
macroscopic-microscopic WS  approach.

This work has  several objectives.
Firstly, we re-examine the old problem of the shell stability of
SHE. Namely, we study  single-particle
spectra resulting from the  (unusually strong)
self-consistent interplay between the nuclear and
Coulomb parts of the Hamiltonian, discuss the major spherical gaps,
and calculate ground-state shapes. Secondly, we analyze the
impact  of the self-consistent treatment on  lifetimes of SHE
and,
in selected cases,  make estimates of fission barriers.
Last but not least, we systematize predicted
 binding energies, separation energies,
and $\alpha$-decay half-lives
 and discuss their dependence on the
choice of the Skyrme parametrization.

The paper is organized as follows.
The models employed are described in Sec.~\ref{models}.
Section \ref{sp_shell} contains the analysis of the single-particle
structure of SHE. Ground-state quadrupole and hexadecapole deformations
and shell corrections
are discussed in
Secs.~\ref{defs} and
\ref{shellc}, respectively.
Binding-energy differences
(i.e., separation energies and $Q_\alpha$ values)
 are analyzed in Sec.~\ref{S2p}.
Section~\ref{alpha} deals with  the particle-decay lifetimes while
Sec.~\ref{fission} contains selected results of  fission
barrier calculations.
Finally, conclusions are contained in Sec.~\ref{conclusions}.

\section{The models}\label{models}

The calculations presented in this paper are
 based on the mean-field theory. The self-consistent results
 have been obtained  using the
Skyrme-Hartree-Fock+BCS method
and
the macroscopic-microscopic results  with
 the deformed WS model. These models
are described in   Secs.~\ref{HFS} and  \ref{NSC}, respectively.

\subsection{Hartree-Fock calculations}\label{HFS}

The Hartree-Fock
 equations have been solved according to the method of
 Ref.~\cite{[Bon85]}. The main advantage  of this
method is that instead of
expanding  single-particle wave functions on an
oscillator basis, one
discretizes them on a three-dimensional mesh, with equally spaced points
along three cartesian axes. This method has been shown
to have an accuracy nearly independent of the
nuclear deformation, and it
does not require an additional basis optimization.
The high efficiency of the method can be attributed  to the
iterative procedure used to solve the HF equations, the
imaginary time step method. One of the features of this method
is that
only those single-particle wave functions which contribute
to the mean field are calculated. In practice, in the following
we have considered all the wave functions which have
a BCS  occupation larger than 10$^{-5}$, i.e., approximately
160 single-particle states for the
protons and 280 for the neutrons. The actual mesh
consisted of 15$^3$ points
 with a mesh step of 1.0~fm.
Under such conditions, the accuracy of  calculations
can be estimated to be
equivalent to that obtained with an   oscillator basis  expansion
in optimized 18 deformed shells.

In the calculations we imposed
 the existence of three symmetry planes.
The resulting shapes are reflection-symmetric, but they can be triaxial.
These assumed self-consistent symmetries are expected to be well justified
for deformations slightly
 beyond the second minimum.
For larger elongations, however, one expects the presence of
odd-multipolarity moments, disregarded in this work.

In the particle-hole (p-h) channel,
the Skyrme effective interactions have been used. The two parametrizations
employed here,
SkP\cite{[Dob84],[Dob95c]} and
SLy7\cite{[Cha95],[Cha95a]},
satisfactorily describe  the
properties of exotic nuclei far from the beta stability line.
Moreover, these interactions have been adjusted to
reproduce  long
isotopic sequences; hence one  can expect them to
have good isospin properties.
SkP has been especially designed to act both in the p-h and
particle-particle (p-p) channels. It
 has an effective mass equal to the free nucleon mass
 ($m^*/m$=1). Therefore, it reproduces properly the single-particle
 level density around the Fermi surface.
 When fitting SLy7,
special attention was  paid
to its ability to describe the properties of
symmetric and asymmetric infinite nuclear
matter. For this force the
 effective mass is rather low,
 $m^*/m$=0.67.
 Surface properties of SLy7 are close
to those of the  SkM$^*$ parametrization. Consequently,
SLy7 is expected to describe
correctly heavy nuclei at large deformations,
in particular fission barriers.
During  the adjustment of the parametrization SLy7, both the tensor
contribution to the spin-orbit term and the two-body center-of-mass
correction have been included self-consistently on the
mean-field level. In our study, however, these terms are treated
perturbatively, i.e., the corresponding corrections to the mean fields
are not taken into account and only added to the
total energy after the self-consistent solutions are found.
Such a procedure may
slightly affect the single-particle levels obtained
in the HF+SLy7 variant discussed in Sec.~\ref{sp_shell}.

The nuclear matter characteristics of SkP and SLy7 are
displayed in Table~\ref{SkPL}. These forces differ mainly by
different values of the isoscalar effective mass and enhancement factor
(i.e., isovector effective mass).
The remaining nuclear matter properties are very similar.
In particular, the corresponding semiclassical
surface energies\cite{[Cam80a]} are almost
identical and equal to that for the SkM* interaction \cite{[Bar82]}
$a_{\rm surf}$=21.8\,MeV, where
it has been adjusted to give correct fission barrier properties.
More refined methods\cite{[Tre86]} to calculate $a_{\rm surf}$
give slightly different numbers which, however, are again very
similar for the three forces SkM*, SkP, and SLy7.
Nevertheless, the fission barriers obtained for SkP and SLy7
are significantly different, Sec.~\ref{fission}. This suggests that the
nuclear matter indicators are too crude a handle
on the results of actual HF calculations at large deformations
where a detailed shell structure is more important than the overall
semiclassical properties.
By using
the two  Skyrme parametrizations, SkP and SLy7, which
differ in their effective mass properties,
we aim at investigating
 the uncertainity in predictions related to the p-h interaction channel.

In the p-p channel, we use a
density-independent
contact  interaction,
   \be\label{DIDI}
   V_\tau^{\delta}(\bbox{r},\bbox{r}')
        = V_{0,\tau} \delta(\bbox{r}-\bbox{r}')~~~~(\tau=\text{n,p}),
   \ee
 for {\em both} Skryme parametrizations employed.
In order to minimize  unphysical fluctuations of the pairing field
in regions of  low level density,
an approximate particle number projection was implemented by
means   of   the   Lipkin-Nogami (LN) 
method\ \cite{[Nog64],[Pra73]}.

 In order to adjust the pairing strengths $V_{0,\tau}$,
 we have estimated an average
  $A$-dependence of the experimental pairing gaps for even-even
 nuclei around $^{254}$Fm.
 As a result, the following expressions for the
 proton and neutron average pairing gaps have been obtained:
 \begin{equation}\label{avD}
     \bar{\Delta}_\tau =  \delta_\tau/A^{1/2},
 \end{equation}
where $\delta_n$=10.304\,MeV and
$\delta_p$=13.072\,MeV.
 The pairing strengths of the interaction (\ref{DIDI})
   have then been
 adjusted to reproduce
the average proton and neutron pairing gaps
 in  $^{254}$Fm.

 Table \ref{T1}  displays the resulting pairing strengths,
  together with  the predicted and experimental pairing gaps in
 $^{254}$Fm. The experimental pairing gaps,
 $\Delta^{\rm exp}$,
  were extracted from odd-even mass differences
 taken from
 Ref.~\cite{[Aud95]}. The quantity
 $\Delta^{\rm LN}+\lambda_{2}$ represents the ``effective" pairing
 gap in the LN approximation (see, e.g., Ref.~\cite{[Mol92b]}).
 The rather different values of $V_0$
obtained in SkP and SLy7  reflect the differences
in their respective  effective masses. Indeed, the larger values
of $V_0$ in SLy7 compensate for the reduced single-particle
level density around the Fermi level due to the low
value of $m^*$ \cite{[Jen86]}.
As shown in Table~\ref{T2},   the  optimized values of
 $V_{0,\tau}$  reproduce rather well
 the average pairing gaps in even-even nuclei
 in the Th-Fm region.

It is well known \cite{[Dob84],[Naz95]}
that traditional models of pairing correlations (e.g., BCS)
 become inappropriate  when approaching
particle drip lines. The main drawback is
the scattering of nucleons from bound shell-model
orbitals to unbound states,  giving rise to an unphysical
component in the nucleonic density with the wrong asymptotic behavior.
This problem becomes particularly severe for neutron-rich nuclei
where the particle continuum lies very low in energy.
On the proton-rich side, which is approached in the present calculations,
 the effect of the continuum is
weaker \cite{[Dob94]}
because of the Coulomb barrier which tends to localize the proton
density in the nuclear interior.

\subsection{Woods-Saxon calculations}\label{NSC}

The
macroscopic-microscopic  method
 is an approximation to the HF approach \cite{[Str67],[Bra72]}.
Its main assumption  is that the total energy
of a nucleus can be decomposed  in two parts,
\be\label{Eshell1}
E = E_{\rm macro}+E_{\rm micro},
\ee
where $E_{\rm macro}$
 is the macroscopic energy
and $E_{\rm micro}$ is the microscopic energy (shell correction)
calculated from a non-self-consistent average deformed potential.

In our study, $E_{\rm macro}$
was  calculated using the  Yukawa-plus-exponential  mass
formula  of
Ref.~\cite{[Mol88]} with parameters of
Ref.~\cite{[Cwi94a]}.
That is,
 similar to Ref. \cite{[Pat91]}, we adjusted the
volume-asymmetry parameter, $\kappa_v$, and the charge-asymmetry parameter,
$c_a$, to obtain good local agreement with known
masses in the heaviest element region.
The optimal values, $\kappa_v$=1.92552 and $c_a$=0.14505 MeV, turned out
to be close
 to the original values of Ref. \cite{[Mol88]}, i.e.,
$\kappa_v$=1.911 and $c_a$=0.145 MeV.

The   shell   correction $E_{\rm micro}$  was   computed
using    the
axially-deformed single-particle Woods-Saxon (WS)
Hamiltonian \cite{[Cwi87],[Ben89]}
with the parameters of Ref. \cite{[Dud81]}.
In the WS calculations
the nuclear surface is defined by means of standard $\beta_\lambda$
axial deformation
parameters:
 \be\label{radius}
R(\Omega ) = c(\beta )R_0\left[
1 + \sum_{\lambda=2,4,6} \beta_{\lambda}Y_{\lambda 0}
(\Omega )\right]
 \ee
with $c(\beta )$ being determined from the volume-conservation condition and
$R_0=r_0 A^{1/3}$.
 The shell correction was calculated according to
the  prescription given in Ref. \cite{[Bra72]}, assuming
the smoothing parameter  of 49.2/$A^{1/3}$ MeV and  a
sixth-order curvature correction.
All  single-particle states lying below a cut-off
energy of 131.2/$A^{1/3}$ MeV above the Fermi level were included.

The pairing energy was calculated  in the LN method using a
state-independent monopole-pairing
interaction.
The pairing strengths and the
 average pairing energy were taken according to Ref. \cite{[Mol92b]}.
The inclusion of the average LN pairing energy led
to a renormalization of the mass-independent constant
in the macroscopic energy formula
\cite{[Mol81],[Mol88]}, $c_0$=1.59741 MeV.
Finally,  the zero-point energy correction
of Ref. \cite{[Pat91]} was added to the total binding energy.

\section{Single-particle structure of SHE}\label{sp_shell}

The shell structure of SHE is governed by several factors, the most
important one being the  large mass  and large electric charge
of a superheavy nucleus.
With increasing $A$, the single-particle spectrum becomes more compressed,
and the binding energy increases
due to the increased range (radius) of the average potential. Moreover,
the orbitals with $\ell$=$\cal N$, $j$=$\ell$$-$1/2, such as $1h_{9/2}$,
$1i_{11/2}$, $1j_{13/2}$, etc., are considerably shifted down in energy
due to a weakened spin-orbit splitting at large values of $A$
(see Refs. \cite{[Sob66],[Mel67]} and Fig.\ 2-30 of Ref. \cite{[Boh69]}).
Indeed, according to the estimates of  Refs. \cite{[Boh69],[Mai95]}, the
one-body coupling constant of the
spin-orbit term  decreases faster
than expected from the
usual radial $A^{-1/3}$-scaling law \cite{[Nil78]}.
For
protons, an additional change in the shell structure comes from the Coulomb
potential which decreases the binding energy of the
single-proton orbitals with small  values of $\langle r^2 \rangle$.
Consequently, the Coulomb potential gives rise to a
lowering of the unique-parity shell
($\ell$=$\cal N$)
 with respect to the normal-parity orbitals with $\cal N'$=$\cal N$$-$1.

Some of the
systematic changes in the single-particle shell structure
are illustrated in Figs.~\ref{espPb} and \ref{esp126A} which display
the spherical WS single-proton energies
in  $^{208}$Pb and
$^{310}126_{184}$, respectively,
 as functions of the Coulomb potential strength $x_{\rm Coul}$.
The latter parameter, varied in the range
0$\leq$$x_{\rm Coul}$$\leq$1,
 renormalizes the electric charge $Z$$-$1 that defines
the strength of the Coulomb potential.
Namely, for $x_{\rm Coul}$=0 the Coulomb term vanishes,  and
for $x_{\rm Coul}$=1 the Coulomb strength  takes its normal value.
In order not to expand  the energy scale too much, the single-particle
levels are shifted in Figs.\ \ref{espPb} and \ref{esp126A} by
\be\label{eshift}
\Delta e = V_{\rm Coul}(r=0)\left[1 - x_{\rm Coul}\right],
\ee
where $V_{\rm Coul}(r=0)$ is the Coulomb energy at  the
center of the nucleus.

The influence of the Coulomb term is  apparent
in Fig.~\ref{espPb}. With increasing $x_{\rm Coul}$,
  high-$j$ shells (most notably: unique parity shells)
  are clearly shifted down
with respect to low-$j$  orbitals. This shift, however,
does not lead to any significant change in the spherical shell structure.
In particular, the Coulomb term does not change the position of magic gaps
50, 82, and 126. The situation is different in the
superheavy nucleus $^{310}$126 (Fig.~\ref{esp126A}). Here, the
Coulomb energy acting together with the energy
shifts due to (i)
the increased
radius and (ii) the change in the
spin-orbit splitting discussed above, induces significant changes
in the shell structure. In particular, the
lowering of the
$1h_{9/2}$ and $1i_{11/2}$ orbitals gives rise to the
(almost complete)  closing
of the spherical gaps at $Z$=82 and 126, and
the appearance of the subshell
closure at $Z$=114 (cf.\ also Fig.\ 2 of Ref. \cite{[Mel67]}
and discussion in Ref. \cite{[Tan79]}).

The position of the magic gaps in the superheavy region
has been discussed extensively in the past, both in
non-self-consistent
\cite{[Cwi92],[Mol92a],[Bra72],[Sob66],[Mel67],[Nil78],%
[Tan79],[Mye66],[Won66],%
[Gus67],[Ros68],[Mos69],[Nil69],[Muz69],[Sob71],%
[Joh72],[Nem72a],[Cha74],[Luk75],[Won76],[Pet76],%
[Cha78a],[Dud78],[Mos78],[Muk80]}
and self-consistent
\cite{[Vau70],[Dav71],[Koh71],[Bas72],[Vau72],[Sau72],[Rou72],%
[Lom73],[Flo73b],[Bei74],[Bra74a],[Cus76],%
[Rou77],[Val77],[Kol77],[Bra78],%
[Koh78],[Ton78],[Ton80],[Gam90],[Boe93],[Ber96]} models.
Early  calculations based on
finite  average potentials
\cite{[Sob66],[Mel67]} and the modified oscillator potential
\cite{[Gus67],[Mos69]}
found $Z$=114 (but not $Z$=126)
and $N$=184 to be the  magic numbers in the superheavy region.
(Similar conclusions were reached in the following works
\cite{[Tan79],[Joh72],[Cha74],[Muk80],[Mol76]}.)
The  magic character of the $Z$=114 gap was questioned by several
authors \cite{[Ros68]} who
pointed out that its size
is sensitive to the details of extrapolation of model parameters;
in particular it
is influenced by the magnitude of the $2f_{7/2}-2f_{5/2}$
spin-orbit splitting
\cite{[Nem72a],[Pet78]}.
The size of the $Z$=114 and $N$=184 gaps in the nucleus $^{298}$114,
predicted in various models, is displayed in Table~\ref{T114}.
In almost all cases, the size of the $Z$=114 gap reflects the
$2f_{7/2}-2f_{5/2}$ proton spin-orbit splitting. In some calculations, e.g.,
HF+SLy7, HF+SkP, and
Br\"uckner-Hartree-Fock (BHF),
the $1i_{13/2}$ proton orbital
appears {\em above} the $2f_{7/2}$ shell. (In several other models,
including WS, HF+SIII
these two shells are practically degenerate,
see Fig. \ref{esp126A}; we also note the similarity
between the HF spectra of Refs. \cite{[Rou72],[Val77],[Ton78]} and
our HF+SkP spectrum.)

Generally, calculated shell effects in the SHE (hence their lifetimes)
depend very strongly on the symmetry dependence [($N$$-$$Z$)/$A$ - term]
of the spin-orbit potential \cite{[Tan79],[Mos78]}. In this context, it is
interesting to compare in Table~\ref{T114}
the Skyrme-HF results with those obtained in the
RMF theory \cite{[Gam90],[Boe93]},
 where the the isospin dependence of the spin-orbit
term is different.

Figures~\ref{esn114}-\ref{esp126}
display the spherical single-particle levels
in $^{294}$114 and $^{310}$126
obtained
in the HF+SkP, HF+Sly7, and WS calculations.
 For the neutrons, all variants  give a rather consistent scenario:
the first significant spherical shell gap above
 $N$=126 is 184, and
 the next large spherical closure is predicted at $N$=228
(see Fig.~\ref{esn126}). It is to be noted that the
neutron shell structure,
including the absolute shell-gap sizes, is
extremely similar in the HF+SkP and WS variants, while the spectrum
obtained in HF+SLy7 shows a characteristic stretching. This can be
attributed
to a fairly small effective mass in SLy7; a smaller
effective mass gives rise to a reduced single-particle level density,
hence to  an overestimation of shell effects
\cite{[Dob95c],[Bra78],[Ton80]}.
For this reason,
 the SkP spectrum (effective mass equal to one)
is close to the WS spectrum.

For the protons (Figs.~\ref{esp114} and \ref{esp126})
the results show strong model dependence.
Namely, the HF+SkP variant  yields the magic proton gap
at $Z$=126; in
the HF+SLy7 variant  the sizes of the $Z$=114 and $Z$=126 gaps are comparable
(although the level density around the $Z$=126 gap is much lower), and
there is a strong preference for the $Z$=114 gap in the WS calculations.
Again,  the SkP interaction predicts smaller
 gaps than SLy7.

In light of a fairly good agreement for neutrons,
the visible discrepancy between the WS results
and  the self-consistent results for protons
might look surprising. This difference, however, has a simple
explanation in terms of surface properties.
Indeed,
it has been argued for some time that a serious
source of uncertainty, when extrapolating the parameters
of average potentials such as WS to the region of SHE, is
the surface diffuseness or surface thickness
\cite{[Mel67],[Nil78],[Tan79],[Dud78],[Lom73],[Val77],[And77]}.
Since self-consistent models
indicate that the
surface thickness parameter for protons is fairly close to that
for neutrons (cf. discussion in Refs. \cite{[Ton80],[Gue80]}
based on the semiclassical extended Thomas-Fermi (ETF) approach),
it has been argued \cite{[Dud78],[And77],[Yar76]} that the proton
diffuseness parameter in the WS model should be systematically increased
in nuclei with large values of $Z$,
in order to compensate for the reduction
of the surface thickness due to the Coulomb potential.
In particular, it has been shown  in Refs. \cite{[Dud78],[And77]} that
the relative size of the $Z$=114 and 126 gaps is
sensitive to the
diffuseness in the proton
average  potential (the increased diffuseness gives
rise to increased $Z$=126 shell gap).
The difference between the HF and WS proton spectra
seen in Figs.~\ref{esp114} and \ref{esp126} can thus be attributed to
a non-self-consistent treatment of surface properties
in the WS model.

In the HFB+Gogny calculations of Ref.~\cite{[Ber96]}
the $Z$=114 gap is small, around 1.4\,MeV, and the
$Z$=126 gap is even smaller (the
spherical $Z$=138 gap is predicted to be around 3\,MeV).
For the neutrons, the
$N$=184 and 228 shells are similar as in our calculations.

Summarizing this section, our self-consistent calculations,
based on the the Skyrme interactions SkP and SLy7,
 favor
the $Z$=126 spherical shell closure over the $Z$=114 gap.
This observation is consistent with the recent results of
the experimental systematics of $B(E2)$ rates \cite{[Zam95]} according
to which $Z$=126 is presumably
the dominant proton shell closure when the global
properties of nuclei in the trans-actinide region
are parametrized in terms of numbers of valence particles,
which depend on the shell closure beyond the experimentally
accessible range.

\section{Equilibrium shapes}\label{defs}

It has been early noted
\cite{[Mye66],[Nil69],[Bol71],[Bol72a]}
that the very
existence of SHE elements is entirely due to shell effects. Indeed,
for $Z$$\gtrsim$100, the macroscopic energy becomes unstable
to shape deformations due to the Coulomb repulsion.
This instability is illustrated in Fig.~\ref{LD} which displays
 $E_{\rm macro}$
calculated using the  diffused-surface Yukawa-plus-exponential  mass
formula (Sec.~\ref{NSC})
as a function of $\beta_2$ for three elements
with $Z$=108, 114, and 126. It is interesting to note that
while for $Z$=108 and  114
the macroscopic energy is unstable to prolate distortions,
for still heavier elements there develops an instability
with respect to oblate deformations.
(In reality, this apparent minimum is  a saddle point,
unstable to triaxial distortions \cite{[Swi74],[Ben75b]}.)
 This softness against oblate
deformations  explains
large oblate ground-state deformations
in nuclei with $Z$$\gtrsim$120 obtained in some  calculations (see below).
As shown in Refs. \cite{[Cwi92],[Ben75b]}, similar results are obtained with
the standard sharp-surface
liquid-drop formula \cite{[Mye67]}; the latter one
is more rigid to deformation and gives rise to slightly wider and
higher  fission barriers.

Since the macroscopic energy of the SHE is unstable with respect to
spontaneous fission, it is the magnitude of the shell correction that
governs their lifetimes. Moreover, since
the macroscopic energy
does not favor spherical shapes
(contrary to the situation in lighter elements), the deformed
shell effects play  a very important role in the description
of SHE. Practically any local deformed gap around the Fermi level
can give rise to a local minimum in the potential energy surface
(PES), i.e., to a multitude of coexisting minima. The
detailed
energy balance  between the local  minima (i.e.,
whether a nucleus is spherical, prolate, or oblate)
  is dictated by
the distribution
 of spherical single-particle orbitals (see Sec.~\ref{sp_shell}).
To illustrate this point,
 potential energy surfaces characteristic of
nuclei from this mass region
are shown in
Figs.~\ref{landscape_WS} (WS, $^{310}$126$_{184}$) and
\ref{landscape} (HF+SLy7, $^{288}$112$_{176}$).
As one can see, in both cases
the pattern of PES is very irregular; it reflects
shell fluctuations in $E_{\rm micro}$.
According to  HF+SLy7,
 the ground state minimum of $^{288}$112$_{176}$ corresponds to
a deformed prolate shape with $Q_{20}$=15\,b ($\beta_2$$\approx$0.11).
There also appear a shallow oblate minimum with $Q_{20}$$\approx$$-$25\,b
which, as we shall see below,
corresponds to  a saddle point (i.e., maximum in the
$\gamma$-direction \cite{[Nil69],[Ben75b]}),
and  a low-lying superdeformed (SD)
minimum at $Q_{20}$$\approx$80\,b. The nucleus $^{310}$126$_{184}$
is predicted to be oblate
($\beta_2$=$-$0.2)
by the WS calculation; the
spherical minimum is about 1\,MeV higher in energy.
Two excited local oblate pockets appear at $\beta_2$=$-$0.4 and $-$0.6
in the region
of  very flat behavior of $E_{\rm macro}$ on the oblate side.
There is also a local well-deformed prolate pocket
($\beta_2$=0.45) beyond the first barrier on the way to fission.
A similar complex pattern of PESs for SHE
has been obtained in previous WS calculations \cite{[Cwi92]}
and in the recent HFB+Gogny calculations of Ref. \cite{[Ber96]}.

\subsection{Axial shapes}\label{axial}

 In self-consistent calculations,
 the shape of a nucleus is characterized
 by expectation
 values of multipole operators rather than by shape deformations.
In our axial HF calculations the
expectation values of multipole moments
\be\label{moments}
\hat{Q}_{\lambda0} \equiv 2 r^\lambda P_{\lambda}(\cos\theta)
\ee
are determined for each self-consistent solution.
For the sake of comparing the HF and WS calculations, one can
introduce  the equivalent deformation parameters, i.e.,
such that the uniform distribution of matter inside the nuclear surface
defined by Eq.~(\ref{radius}) reproduces the HF expectation
values $Q_{\lambda0}$=$\langle\hat{Q}_{\lambda0}\rangle$ of
multipole moments (\ref{moments}).
By keeping  the second-order
terms
in $\beta_2$ and $\beta_4$,
the expressions for the mass quadrupole and hexadecapole
moments, which define the equivalent deformation parameters, are
(cf.\ Ref.\cite{[Has88]})
\begin{eqnarray}\label{defs1}
   Q_{20}&=&\frac{3}{\sqrt{5\pi}} A R_0^2
          \left( \beta_2 + \frac{2}{7}\sqrt{\frac{5}{\pi}}\beta_2^2 +
            \frac{20}{77}\sqrt{\frac{5}{\pi}}\beta_4^2 +
            \frac{12}{7\sqrt{\pi}}\beta_2\beta_4 \right),\nonumber \\
   Q_{40}&=&  \frac{1}{\sqrt{\pi}} A R_{0}^{4}
          \left( \beta_{4} + \frac{9}{7\sqrt{\pi}}\beta_{2}^{2} +
            \frac{729}{1001\sqrt{\pi}}\beta_{4}^{2} +
            \frac{300}{77\sqrt{5\pi}}\beta_{2}\beta_{4} \right),
 \end{eqnarray}
where  $R_{0}=1.2 A^{1/3}$. In a similar way one can define the relation
between neutron (proton) moments and
the corresponding deformation parameters.

Figure~\ref{b27} displays, in the form of contour maps,
 the ground-state equilibrium values
of the equivalent $\beta_2$  obtained in the
HF+SkP and HF+SLy7 variants. The general pattern
is very similar for both Skyrme interactions. Namely,
for $N$$\lesssim$176
all the nuclei considered are predicted to be prolate,
and they are spherical for $N$$\gtrsim$180. These two regions are
separated
by a narrow band of transitional nuclei with
 small equilibrium deformations.
The largest deformations, around $\beta_2$=0.26,  are
 obtained for the lightest nuclei considered, i.e.,
 for $Z$$\lesssim$114 and $N$$\lesssim$162.
The stabilizing role of deformation
in the region of the known heaviest elements
is well known
\cite{[Cwi83],[Cwi85],[Bon86a],[Sob87],[Sob89],%
[Pat89],[Cwi92],[Laz94],[Cwi94a],[Mol94a],[Smo95],[Mue88]};
it has been attributed to the deformed shell gaps
at $N$=162 and $Z$=108. Similar ground-state deformations of
the heaviest elements with $Z$$\leq$114 have recently
been obtained in the HF+SIII calculations \cite{[Taj96]}.

Figure \ref{b47}
shows the ground-state equilibrium
values of the equivalent $\beta_4$ deformations.
Again, the isotopic and isotonic
dependencies of $\beta_4$ are
very similar for both effective interactions.
All nuclei which are deformed in the ground state
have negative values of $\beta_4$.
The largest hexadecapole distortions,
$\beta_4$$\approx$$-$0.10,  are
calculated in nuclei around $Z$=112 and $N$=168.

The ground-state quadrupole deformations obtained in the  WS 
model are
displayed in Fig.~\ref{shapes}. For $Z$$\lesssim$118, the results are
close to those calculated in  the
self-consistent  models. The significant differences
are obtained for nuclei around $Z$=122 and $N$=166 (WS
predicts very deformed oblate shapes with $\beta_2$$\approx$$-$0.4),
and for the isotopes with $Z$$>$124 and those around
$Z$=122 and $N$=190 where
WS  yields deformed oblate shapes with $\beta_2$$\approx$$-$0.2.
In most cases, our present WS results
are very close to the WS results of Ref.~\cite{[Cwi92]}
(cf. Figs. 2 and 3 therein).

In order to relate our results to previous work,
we present in Table~\ref{T3} the quadrupole
and hexadecapole ground-state deformations obtained in the WS model,
finite-range droplet model (FRDM) of Ref. \cite{[Mol95]},
extended Thomas-Fermi with Strutinsky-integral model (ETFSI)
\cite{[Abo95]}, and in our HF+SLy7 and HF+SkP calculations. For the
lightest systems, $^{264}$108 and $^{270}$110,
all calculations agree with one another.
The results of both HF variants
are very close for all nuclei considered.

As illustrated in Figs.~\ref{landscape_WS} and \ref{landscape}
and discussed above,
 one expects  competition between
near-lying prolate, oblate, and  SD shapes. Hence,
 depending on those model details which can influence the magnitude
of shell correction,
the calculated
ground state of a superheavy nucleus can be prolate, oblate, or  SD.
This is clearly seen in Table~\ref{T3}, especially
 for the heavier isotopes with
$Z$$>$110.

In the FRDM calculations,
 the oblate minimum is favored
for weakly deformed nuclei with $N$$\approx$174.
A similar result was obtained in a recent HFB+Gogny study
of Ref.~\cite{[Ber96]}.
 A SD ground state ($\beta_2$$\gtrsim$0.5)
is predicted in the FRDM
for the very heavy systems with $Z$$\geq$118 and $N$$\gtrsim$194.
In the
ETFSI model, many nuclei
(practically all isotopes with $Z$$\geq$112)
have very elongated ground states with
$\beta_2$$\gtrsim$0.4
(cf. Table~\ref{T3}).
The HFB calculations of
Ref.~\cite{[Ber96]}
 are in good agreement with our results.
Except for the few isotopes that are
 predicted to be  oblate in the HFB+Gogny calculations,
 the general features of the potential energy
surfaces  are similar.

The transition from deformed to spherical shapes
predicted for $N$$\gtrsim$180
can be associated with the spherical neutron shell closure at
$N$=184 (see Sec.~\ref{sp_shell}).
 The effect of the $Z$=126 spherical proton gap
appears to be weakened as compared to the known nuclei.
The reason for this is twofold. Firstly,
as discussed in Sec.~\ref{sp_shell} above, the gap itself is significantly
quenched in the SHE. Secondly, unlike in the lighter
elements,  spherical shell effects {\em are not} further stabilized by the
liquid-drop energy contribution; i.e.,
for $Z$=126
the macroscopic energy has
a maximum at the spherical shape. Consequently, the spherical
proton gap at $Z$=126 cannot be considered as truly
magic: the neutron-deficient
$Z$=126 isotopes are all predicted to be deformed.

\subsection{Triaxial shapes}

In order to illuminate  the role  of triaxial degrees of freedom,
in selected cases
we performed the
constrained HF calculations in the ($Q_{20}, Q_{22}$) plane.
Figures~\ref{qg112} and~\ref{qg126} display the self-consistent PES's
calculated for the HF+SLy7 interaction and
for
$^{288}$112 and $^{310}$126, respectively.
Here, the triaxial deformation $\gamma$ is defined as:
\begin{equation}
\tan{\gamma}=\sqrt{2}\frac{Q_{22}}{Q_{20}}.
\end{equation}

As discussed in Sec.~\ref{axial},
the nucleus $^{288}$112 is weakly prolate-deformed
in its ground state. The secondary oblate minimum seen
in Fig.~\ref{landscape} corresponds to a saddle point.
The first fission barrier represents  an axial saddle point
at $Q_{20}$=42\,b.
 A secondary SD minimum,
at $Q_{20}$=80\,b,
 can be seen at the edge of the map and is
 more clearly displayed in Fig.~\ref{landscape}.
In short, the static properties
of the nucleus $^{288}$112 can be well described
by calculations constrained to axial shapes.

The situation is rather different in $^{184}$126. For both interactions,
  the ground state is spherical and there is no indication
for the oblate minima predicted in WS.
The static path to fission
goes through triaxial saddle points. There is
only one saddle point at $\gamma$$\approx$$-$30$^\circ$ for SkP
and two triaxial saddle points and a secondary triaxial minimum
for SLy7.
Clearly, the microscopic study of the fission
process in $^{184}$126 would require a  dynamical calculation,
fully exploring the triaxial degrees of freedom.

\section{Shell corrections}\label{shellc}

To compare relative strengths of shell effects in different
models, it is convenient to introduce  the shell  energy,
$E_{\rm shell}$. In WS, the shell energy is defined as
the difference between the total binding energy and the spherical part
of the macroscopic energy
\be\label{Eshell}
           E_{\rm shell}=E-E_{\rm macro}({\rm spherical})
   = E_{\rm micro}({\rm spherical}) + E_{\rm def},
\ee
i.e., it is the sum of the spherical microscopic energy and
the total deformation energy $E_{\rm def}$.
In order to extract $E_{\rm shell}$ from the HF calculations,
we applied  a very similar procedure. That is, from the
self-consistent ground-state
energy we substracted the same $E_{\rm macro}({\rm spherical})$ as in the
WS calculations.
In addition, the obtained HF values of $E_{\rm shell}$
were shifted by a (small) constant in such a way that their values at
the $Z$=108, $N$=162 nucleus become
equal to the WS value. This procedure aims at comparing the
shell effects calculated in different models irrespectively of a
generic smooth $Z$ and $N$ dependence and a constant offset.
(Another way
of extracting $E_{\rm shell}$ from the HF calculations would be to perform
the Strutinsky renormalization procedure based on the HF single-particle
energies \cite{[Bra75]}.
However, because of the unphysical description of
the positive-energy discretized continuum, especially
for the proton-rich SHE,  the standard renormalization technique
is not very useful \cite{[Naz94]} and we did not pursue this
approach further.)

The contour maps of $E_{\rm shell}$ are displayed  in Fig.~\ref{esh}.
Both HF calculations clearly show the presence
of strong shell stability around the ``doubly-magic"
nucleus  $^{310}126$.
This result markedly differs from  both that of the WS variant, where
the island of shell stability is concentrated around
the nucleus $^{296}$114,
and the result of the FY
model calculations by
M\"oller and Nix \cite{[Mol92a]} who  obtained the
maximum shell energy
at  $^{292}$114.

The position of the island of shell
stability
does influence the mass differences that largely determine
proton and neutron separation energies (Sec.~\ref{S2p}) and
particle emission probabilities, and hence
the partial lifetimes (see Sec.
\ref{alpha}).
The fact that the values of $E_{\rm shell}$ have very different ranges
in HF and WS calculations (Fig.~\ref{esh}) implies that
the $Z$ and $N$ dependence of the total energy also differs; although
both WS and HF calculations  reproduce  properties of known heavy elements,
they {\em extrapolate very differently} with
the particle numbers.

\section{Separation and $\alpha$-decay energies}\label{S2p}

Based on the calculated values of binding energies (or mass excesses,
$M_{\rm exc}$), separation and nuclear reaction energies can
be extracted. These quantities involve masses of two or more
nuclei and they are sensitive probes of symmetry properties
of the effective interaction.

\subsection{Two-neutron and two-proton separation energies}

Firstly, we calculate two-neutron and two-proton separation energies
of even-even SHE.
These quantities,  defined in terms of
$M_{\rm exc}$,
\begin{eqnarray}\label{separ}
S_{2n}&=&-M_{\rm exc}(Z,N)+M_{\rm exc}(Z,N-2)+2M_{\rm exc}(0,1),\\
S_{2p}&=&-M_{\rm exc}(Z,N)+M_{\rm exc}(Z-2,N)+2M_{\rm exc}(1,0)
\end{eqnarray}
[$M_{\rm exc}(1,0)$=8.071\,MeV,
$M_{\rm exc}(0,1)$=7.289\,MeV],
 probe the variations in the mass surface in the directions of $N$ and $Z$.

The calculated (HF and WS)
   two-proton and two-neutron separation energies
       are  plotted in Figs.~\ref{s2p} and \ref{s2n}, respectively.
     In spite of rather different values of   $E_{\rm shell}$
     obtained in the self-consistent calculations
and in the macroscopic-microscopic
     approach,
      the behavior of two-particle separation energies
      is  strikingly similar. The
      excellent
      agreement obtained for the known heaviest elements seems to
hold for very exotic superheavy systems as well.

       The
  two-proton drip line, $S_{2p}$=0,
(indicated by a thick solid line in Fig.~\ref{s2p})
  goes very  smoothly between
   the nuclei $^{264}110_{154}$ and $^{314}126_{188}$. In agreement
   with the results presented in Sec.~\ref{sp_shell},
   this  regular behavior of $S_{2p}$ as a function of $N$
  does not  indicate any   strong shell effects.
Interestingly, the nucleus $^{310}126_{184}$, predicted to be strongly
   stabilized by the shell effects in HF,  is calculated
   to be proton-unstable (see Fig.~\ref{s2p}).
   However, due to very large Coulomb barriers,
  nuclei predicted to have proton separation energies $S_p>-1.5$\,MeV
  are practically stable to proton emission, i.e., this decay channel
  can be practically disregarded as compared to other decay modes
  (cf. discussion in Sec.~\ref{alpha}).
A similar conclusion was drawn by
Tondeur \cite{[Ton78]} who predicted proton instability 
and a long proton emission lifetime of $^{310}$126
but  estimated very short $\alpha$- and $\beta^+$-decay half-lives.

 As seen in Fig.~\ref{s2n},
  the isotopes considered are very  far from
  the neutron drip  line. A strong spherical shell effect at $N$=184
  is clearly visible, especially in HF calculations. The effect
  of the deformed
  closure at $N$=162, particularly pronounced in SLy7,
   is also seen.

\subsection{Alpha-decay energies}

As mentioned earlier, the main decay
mode for the known heaviest elements is $\alpha$-decay.
The calculated $Q_\alpha$-values,
\be\label{Qalpha}
Q_\alpha = M_{\rm exc}({\rm Z,N}) -
M_{\rm exc}({\rm Z-2,N-2}) - M_{\rm exc}(2,2)
\ee
[$M_{\rm exc}(2,2)$=2.425 MeV],
are displayed
in Fig.~\ref{qalpha}.
The landscapes of $Q_\alpha$ obtained in the HF+SkP and HF+SLy7 models
are very  similar. For the nuclei around $Z$=112 and  $N$=162,
 the WS variant  yields
$Q_\alpha$-values extremely close to those of HF+SkP.
This is illustrated in Table~\ref{CDATA} which displays calculated
 values of $Q_\alpha$ for the {\em known} $\alpha$-decay
chain of $^{264}$Hs and compares them to the data.
 The obtained good agreement with experiment  and with
the  results of the WS model (which was optimized locally in this region)
is a necessary condition for
trusting predictions of our self-consistent calculations in the region
of unknown SHE.

A noticeable difference is obtained in the region  around
($Z$=116, $N$=184) where  WS  predicts values which
are  $\sim$2\,MeV larger than those in self-consistent calculations.
This shift in $Q_\alpha$ results from
 different predictions regarding  proton
shell effect ($Z$=114 vs. $Z$=126, see Sec.~\ref{sp_shell}) and has
a direct consequence for $\alpha$-decay half-lives  in this region
(see Sec.~\ref{alpha} below).

\subsection{Beta-decay energies}

In the region of  $\beta$-stability, both
$Q_{\beta^{-}}$ and $Q_{\rm EC}$ are negative.
Since our calculations are limited to the even-even nuclei, in
order to estimate 
the beta-decay energies
\begin{eqnarray}
Q_{\beta^{-}}&=&M_{\rm exc}(Z,N)-M_{\rm exc}(Z+1,N-1),  \nonumber \\
Q_{\rm EC} &=& M_{\rm exc}(Z,N)-M_{\rm exc}(Z-1,N+1), \nonumber \\
Q_{\beta^{+}}&=& Q_{\rm EC} -2m_e,
\end{eqnarray}
masses of the odd-odd systems have been obtained in the 
quasi-particle approximation \cite{[Bei75a],[Smo93]}. The 
final result for the beta-decay energy
 of  the parent nucleus ($Z$, $N$)
 can be expressed in terms of binding energies
of neighboring even-even nuclei and the ``effective" LN pairing gaps:
\begin{eqnarray}\label{EC}
Q_{\beta^{-}}& = & \Delta M_{H-n} -\Delta^{\rm LN}_p (Z,N-2)-
                          \lambda_{2,p}(Z ,N-2)- 
            \Delta^{\rm LN}_{n}(Z ,N-2)-
                         \lambda_{2,n}(Z ,N-2) \nonumber \\
   & + &  \frac{1}{4}\left[3B(Z,N )
           -{B}(Z,N-2) -B(Z +2,N )
           -B(Z+2,N -2)\right], \nonumber \\
Q_{\rm EC} & = & - \Delta M_{H-n} -\Delta^{\rm LN}_{p}(Z-2,N )-
                          \lambda_{2,p}(Z -2,N )- 
                \Delta^{\rm LN}_{n}(Z-2,N )-
                         \lambda_{2,n}(Z -2,N ) \nonumber \\
   & + &    \frac{1}{4}\left[3B(Z,N ) -
                {B}(Z-2,N )-{B}(Z-2,N+2)-
                {B}(Z ,N +2)\right],
\end{eqnarray}
where $\Delta M_{H-n} = m_{n}-M_H$,
 $M_{H}$ and $m_{n}$ denote the hydrogen atom
and neutron masses, respectively,
and ${B}$ is the binding energy
of an even-even nucleus.

The predicted  $\beta$-stability valley
is shown in Fig.~\ref{betstab}. According to our
calculations,
the spherical nuclei from the
neighborhood of $^{298}$114 are beta-stable.
On the other hand, the predicted
``doubly-magic"  
 nucleus $^{310}$126
lies outside the stability valley.
This result  agrees with the conclusions of
previous studies  \cite{[Mol94a],[Ton78],[Ber96]}
where the nuclei around $^{310}$126
have also been calculated to be $\beta^+$/EC-unstable.
According to the estimates of Ref.~\cite{[Mol94a]},
Gamow-Teller transitions
are  of the order of 0.3\,s for $Z\approx 128$
and $N\approx 188$. Kumar \cite{[Kum89]} applied
the semi-empirical estimate for
$\beta$-decay and EC half-lives \cite{[Fis72]}
and obtained the value of 18\,s for  $^{310}$126.
Thus, in this region, the $\beta^+$/EC  process
is expected to be much  slower  as
compared to the $\alpha$-decay
\cite{[Ton78]} (see Sec.~\ref{alpha}).
On the other hand,
for the neutron-rich isotones with $N$$\approx$184 the
$\beta^{-}$ emission is expected to 
 compete with  $\alpha$-decay
and fission \cite{[Ton78]}.

\section{Particle decay lifetimes}\label{alpha}

\subsection{Alpha decay half-lives}

In order to estimate
the $\alpha$ decay half-lives, $\rmT_\alpha$,
we have used the phenomenological formula of Viola and
Seaborg \cite{[Vio66]} with the parameters of Ref. \cite{[Pat89a]},
\be\label{TVS}
\log_{10}{\rmT}^{\rm VS}_\alpha(\mbox{sec})
  = (1.66175\,{Z}_p-8.5166)/\sqrt{Q_\alpha} -
(0.20228\,{Z}_p+33.9069),
\ee
where $Z_p$ is the atomic number of a parent nucleus
 and $Q_\alpha$ is in MeV.
The  
$\alpha$-decay half-lives
calculated in the HF+SkP, HF+SLy7, and WS models
 are  shown in Fig.~\ref{Tal}
for the even-even SHE.

For the known heaviest elements, there is fairly good agreement
between all three variants of calculation. In particular,
 our predictions for
nuclei $Z$$\approx$110-114 and $N$$\approx$154-170
agree with those of Refs. \cite{[Smo96]} and \cite{[Ber96]}.
Especially pronounced  in the HF+SLy7 model,  there appears
a region of  increased stability  associated
with the deformed shell closure at $N$=162 \cite{[Cwi83],[Cwi92]}.
Table~\ref{TDATA} displays calculated values of  $\rmT_{\alpha}$
for the nuclei
which are the most promising candidates to be synthesized
in the near future.

The HF+SLy7 model predicts an increased stability
with respect to the $\alpha$-decay also  for nuclei with
 $N$$\approx$172
and $Z$ ranging from 110 to 118.
 It originates
from the deformed  neutron gap that
opens up for $N$=172-174 at large negative  values of $\beta_4$.
The resulting rapid change of the slope of the shell energy
as a function of particle number around $N\approx$170 is
seen in  Fig.~\ref{esh}.

Both HF  calculations differ significantly from the
WS  calculations for the heavier $N$$>$180 isotones
and predict a considerably  smoother decrease of $\rmT_\alpha$
with increasing proton number. This discrepancy can be attributed
to the difference in the location of the spherical
proton shell which is moved up from $Z$=114
 to $Z$=126.
The maximum in $\rmT_\alpha$ at $N$$\approx$182 can be attributed
to the spherical neutron shell effect at $N$=184. 
Because the proton shell effect appears at  higher values of $Z$,
HF calculations predict longer alpha-decay half-lives 
for all nuclei with $N$$\approx$182 as compared to the
present  WS calculations
and the previous   calculations
of Refs.
\cite{[Cwi92],[Mol94a],[Mol92a],[Luk75],[Smo96],[Sob94]}
based on the macroscopic-microscopic approach.
On the other hand we note a similarity
between our  HF+Sly7 results  and
those of Ref.~\cite{[Ber96]} within the HFB-Gogny model.

It should be emphasized that Eq.~(\ref{TVS}),
based on the WKB approximation,  provides a rather
crude estimate of $\rmT_\alpha$ since it disregards  many structure
effects \cite{[Dec94]} such as deformation, configuration
changes, and so on. However, based on the experience with
the known heaviest elements, the phenomenological expression
(\ref{TVS}) usually gives a right order of magnitude for 
$\rmT_\alpha$; for more precise estimates the microscopic calculations
of $\alpha$ decay  are called for.

\subsection{Proton emission half-lives}

Nuclei with negative proton separation energies are 
unstable to proton emission. 
The proton emission half-lives depend primarily
on the proton separation energy and orbital angular momentum but
very weakly on the details of the parameters of the 
corresponding
proton  optical model potential \cite{[Hof89],[Naz96],[Poe96]}.

The proton decay lifetimes have been estimated using
the WKB expression for the partial width:
\begin{equation}\label{WKB}
\Gamma_{p} = \theta^2{\cal N}\frac{\hbar^2}{4\mu}\exp\left[
-2\int_{r_{\rm in}}^{r_{\rm out}}dr k(r)\right],
\end{equation}
where $\theta^2$ is the  proton spectroscopic factor, 
 $\mu$ is the reduced mass,
$r_{\rm in}$ and $r_{\rm out}$  are the classical inner and outer
turning points, respectively, ${\cal N}$ is the normalization factor,
and $k(r)$ is the wave number given by
\begin{equation}\label{kr}
k(r)=\sqrt{\frac{2\mu}{\hbar^2}|Q_{p}-V_{p}(r)|}.
\end{equation}
In Eq.~(\ref{kr}), $Q_{p}$=$-S_{p}$$>$0, and  $V_{p}(r)$ is the average
proton potential (including the Coulomb potential).

In our calculations, the proton potential was approximated by that of
the WS model, and it has been  assumed that $\theta^2$=1 and the
proton orbital angular momentum, $\ell$, is zero. Because of the
neglect of the centrifugal barrier, the calculated proton half-lives,
$T_p$, are, in general, grossly underestimated. That is, the predicted
values of  $T_p$ should be viewed as lower limits of proton half-lives.

The calculations have been performed for the even-even and odd-even
nuclei with   $S_{p}$$<$0 (see  Fig.~\ref{s2p}).
Since the proton separation energy involves the binding energy
 of the odd-even system (Eq.~(\ref{separ})) to calculate the latter
quantity, we employed  the 
quasi-particle approximation (see discussion before
 Eq.~(\ref{EC})). The 
final result for the proton separation energy
 of  the parent nucleus ($Z$, $N$)
 can be expressed in terms of the proton Fermi energy, $\lambda_p$,
 and the ``effective" LN pairing gaps:
\begin{equation}\label{Qp}
Q_{p}= 
\left\{ \begin{array}{ll}
\tilde{\lambda}_p(Z,N)-
       \tilde{\Delta}_{LN}^{p}(Z,N)-
       \tilde{\lambda}_{2p}(Z,N) & ~~\mbox{if $Z$ is even}\\
 \tilde{\lambda}_p(Z,N) +
       \tilde{\Delta}_{LN}^{p}(Z,N) +
       \tilde{\lambda}_{2p}(Z,N)  & ~~\mbox{if $Z$ is odd}
\end{array}
\right.
\end{equation}
Here the tilde  denotes the
average value, e.g.,
\begin{equation}
\tilde{\lambda}_{p}(Z,N) \equiv
\frac{1}{2}\left[\lambda_{p}(Z-2,N)+\lambda_{p}(Z,N)\right].
\end{equation}

The obtained  proton-emission half-lives are displayed 
in Fig.~\ref{tprot}. For a given isotonic chain, 
the values of $\rmT_p$ for odd-$Z$ nuclei are dramatically
lowered with respect to the even-$Z$ neighbors. As seen
in Eq.~(\ref{Qp}), this odd-even
effect is due to the proton  pairing energy.  

Due to large Coulomb barriers in SHE,
nuclei with $Q_{p}$$<$1.5\,MeV
are practically proton-stable.
As seen in  Figs. \ref{Tal} and \ref{tprot}, the $\alpha$-decay
always wins  for all nuclei 
considered.
That is, the contribution 
of proton emission to the total decay
width can be practically neglected.

\section{Fission barriers}\label{fission}

The self-consistent static paths to fission have been computed in
a few selected cases. The results
 for nuclei  $^{288}$112 and
$^{310}$126 
 are displayed in
Figs. \ref{landscape} and \ref{qgo126}, respectively.

The existence of a rather high fission barrier in  $^{310}$126
(about 14 MeV for SLy7
and 16 MeV for SkP)
is a consequence of increased  stability of the
 $Z$=126 and  $N$=184 system. Interestingly, the patterns of PES
obtained in both HF calculations
for  $^{310}$126 are very similar. The increase in the barrier height
for SkP is due to 
the lower energy of the 
 ground-state configuration in this model
(see Fig.~\ref{esh}).
 Calculations based on the macroscopic-microscopic 
method predict the fission barriers to be significantly
 smaller
(cf., e.g., Ref.~\cite{[Cwi92]}),
due to the  absence of the spherical $Z$=126 shell closure
in those models.

{}Fission barriers are reduced by a few MeV
 if the non-axial degrees of freedom
are taken into account (see Figs.\ \ref{qg112} and \ref{qgo126}).
 Nevertheless,
the resulting barriers
are high enough
to suppress the fission process
as compared to the alpha emission.

The heights of axial fission barriers computed in  the HFB+Gogny
model \cite{[Ber96]} are very
similar to our axial results (non-axiality has been ignored
in their calculations).
They found fission half-lives to be of the order of 10$^{6}$
years
for the $N$=184 isotones (with  barriers of the order of 14 MeV).
A
substantial decrease of $T_{\rm sf}$ due to the non-axiality can be expected.
(A rough estimate gives a decrease by almost 8 orders
of magnitude for  SLy7.) In this context, however,
it should be
emphasized that the
height of the fission barrier obtained along the dynamic path
is expected to  be higher than that obtained
along the static path, and the effect of
non-axiality  is reduced in dynamic calculations 
\cite{[Smo95],[Bar81]}.
Rather high fission barriers in SHEs (about 14 MeV) are also predicted
by Tondeur in Ref.~\cite{[Ton78]}.

The SLy7 static fission path for
the nucleus $^{288}$112  displayed
in Fig. \ref{landscape} shows the presence of the
SD minimum which gives rise to a two-humped structure.
(Note that the non-axial degrees of freedom do not influence
the height of barriers in this region (see Fig.~\ref{qg112}).)
Both barriers (inner and outer)
 have approximately the same height of about 6\,MeV.
Because of the strong Coulomb repulsion between the fission
fragments,
the barrier of 6\,MeV is the minimum to ensure
the stability of a SHE. Hence 
the fission process will probably dominate
in  this region.
A similar pattern for the static fission path
has been obtained in Refs.~\cite{[Smo95],[Ber96]}.
In Ref.\ \cite{[Cwi92]}, the inner barrier has been found to
be lower by 2\,MeV compared to our result. Such a
decrease can be attributed to the 
weaker shell effects predicted in \cite{[Cwi92]}.

The self-consistent barrier calculations
performed here for a few nuclei
confirm the general pattern predicted 
in Refs.~\cite{[Cwi92],[Smo95]}.
Namely,  in the region
of the deformed $N$=162 neutron shell,  the $\alpha$-decay
mode   is
favored due to strong shell effects
which hinder the fission process.
On the other hand,
for nuclei with $N$$>$166 and $Z$$\approx$112 the fission mode
dominates. This can be justified, e.g.,  by a
 very steep
increase of the shell energy in the HF+SLy7 calculations leading
to a rapid decrease of the fission barriers. 
 Since our self-consistent
models do not predict an increased stability in the neighborhood
of $Z$=114 and $N$=184, there will probably be  
strong competition 
between spontaneous fission and $\alpha$-decay.
Heavier isotones of $N$=184 are expected to decay mainly via
the $\alpha$  emission as one is approaching
the spherical magic shell $Z$=126. 

The SHEs with $Z$$>$110 were also analyzed by
Kumar \cite{[Kum89]} within the dynamic
deformation model. He obtained  $\alpha$-decay half-lives
much larger than ours. For instance,
his values of $\rmT_\alpha$ for $^{298}$116 and  $^{310}$126
are 10$^{1.85}$\,y and  10$^{10.04}$\,y, respectively. 
According to this model, the nucleus 
$^{298}$114  decays by fission and  $^{298}$116
decays by $\beta^+$/EC.

\section{Conclusions}\label{conclusions}

We have applied the HF method with two different Skyrme
interactions, SkP and SLy7,
and the WS macroscopic-microscopic approach to the
study of  SHEs
 with
108$\leq$$Z$$\leq$128 and 150$\leq$$N$$\leq$192.
The overall agreement with existing experimental data
for the heaviest known elements
 is satisfactory. In particular,
the HF
calculations predict  the increased
stability due to the deformed shell effect at $N$=162.
However, the self-consistent and macroscopic-microscopic
 approaches
give different results when extrapolated to still
 heavier elements.
The fact that the former methods treat  correctly
 the interplay between
the nuclear and Coulomb properties, which is crucial for the 
physics of SHEs,
gives us more  confidence in
self-consistent  results.

Predictions have been made for binding energies, shell
energies, deformations, reaction energies,
and half-lives. In general, both SkP and SLy7
Skyrme parametrizations give similar results.

According to our study, the
doubly-magic superheavy element is that with $Z$=126 and $N$=184.
This system is unstable with respect to the $\beta^+$/EC, proton, and
 $\alpha$ decays, and it 
fissions. However, its half-life is governed by the
$\alpha$-decay 
($\rmT_\alpha$$<$1\,$\mu$s).
The alleged magic gap at $Z$=114 does not show up  in
our HF calculations. Its presence in the models,
based on the macroscopic-microscopic approach,
is due to the lack of a self-consistent treatment of surface
properties (in particular: proton diffuseness). 
In the self-consistent theory we obtain an
 increased shell stability of spherical
nuclei with $N$$\approx$184 and $Z$$>$114, which gives rise
to increased $\alpha$-decay half-lives and opens up
a possibility to access experimentally new
higher-$Z$  superheavy
elements.

\section*{Acknowledgments}
Useful discussions with Sigurd Hofmann are gratefully acknowledged.
The Joint Institute for Heavy Ion
 Research has as member institutions the University of Tennessee,
Vanderbilt University, and the Oak Ridge National Laboratory; it
is supported by the members and by the Department of Energy
through Contract No. DE-FG05-87ER40361 with the University
of Tennessee. 
Oak Ridge National
Laboratory is managed for the U.S. Department of Energy by Lockheed
Martin Energy Research Corp. under Contract No.
DE-AC05-96OR22464.
This research 
 was supported in part by the U.S. Department of
Energy through Contract No. DE-FG05-93ER40770,
by the Polish Committee for
Scientific Research under Contract No.~2~P03B~034~08,
and by the Belgian Office for Scientific Policy under Contract ARC~93/98-166.
Numerical calculations were performed at
The Interdisciplinary Centre for Mathematical and Computational Modeling
(ICM) of Warsaw University, and at the National Energy Research
Supercomputer Center at the Lawrence Berkeley Laboratory.


\newpage

\narrowtext
\begin{table}
\caption[A]
{Nuclear matter characteristics of the SkP and SLy7 Skyrme
parametrizations. Standard definitions of the energy per particle
$E/A$, Fermi momentum $k_F$, incompressibility $K_\infty$,
isoscalar effective mass $m^*/m$, dipole enhancement factor $\kappa$,
and symmetry energy $a_{\rm sym}$ are collected,
e.g.,  in Ref.\ \cite{[Cha95a]}.
The surface energy $a_{\rm surf}$ is calculated 
using  the  expression
given in Ref.\ \cite{[Cam80a]},
Eqs.\ (9-12) for $\beta$=1/18 and $\gamma$=1/3.
}
\begin{tabular}{lccccccc}
   & $E/A$ & $k_F$ & $K_\infty$ & $m^*/m$ & $\kappa$
   & $a_{\rm sym}$ & $a_{\rm surf}$\\
   & (MeV) & (fm$^{-1}$) & (MeV) & & & (MeV) & (MeV) \\ \hline
SkP & $-$15.95 & 1.34 & 201 & 1.00 & 0.35 & 30 & 21.8 \\
SLy7 & $-$15.91 & 1.33 & 230 & 0.69 & 0.25 & 32 & 21.5
\end{tabular}
\label{SkPL}
\end{table}

\narrowtext
\begin{table}
\caption[A]{Strengths of the zero-range pairing interaction,
Eq.~(\protect\ref{DIDI}), adjusted to reproduce the
proton and neutron
average pairing gaps in  $^{254}$Fm in the
SkP and SLy7 variants of calculation.
The resulting ``effective" odd-even mass difference,
$\Delta^{\rm LN}+\lambda_2$, is compared to the average gap,
Eq.~(\protect\ref{avD}), and the experimental value
extracted from the odd-even mass
differences \protect\cite{[Aud95]}.
All quantities are in MeV.}
\begin{tabular}{lcccccc}
  & \multicolumn{2}{c}{SkP} & \multicolumn{2}{c}{SLy7} & & \\
    & $V_{0}$ & $\Delta^{\rm LN}+\lambda_{2}$ &
      $V_{0}$ & $\Delta^{\rm LN}+\lambda_{2}$ &
      $\bar{\Delta}$ & $\Delta^{\rm exp}$  \\ \hline
 protons &   245    & 0.823  & 290 & 0.817 &  0.820 &   0.768   \\
 neutrons &   190   & 0.645  & 250 & 0.642 & 0.647 &   0.645
\end{tabular}
\label{T1}
\end{table}

\narrowtext
\begin{table}
\caption[A]{Odd-even mass differences
 $\Delta^{\rm LN}+\lambda_2$ calculated
  in the
 HF+SkP+LN approach
compared to the average gaps,
Eq.~(\protect\ref{avD}), and the experimental odd-even
 mass differences
\protect\cite{[Aud95]} for several actinide nuclei.
All quantities are in MeV.}
\begin{tabular}{ccccccc}
  & \multicolumn{3}{c}{neutrons} & \multicolumn{3}{c}{protons}  \\
 Nucleus    & $\Delta^{\rm LN}+\lambda_{2}$ &
      $\bar{\Delta}$ & $\Delta^{\rm exp}$
       & $\Delta^{\rm LN}+\lambda_{2}$ &
      $\bar{\Delta}$ & $\Delta^{\rm exp}$
        \\ \hline
 $^{232}$Th  &   0.65    & 0.68  & 0.75  & 0.89 &     0.86 & 1.03  \\
 $^{236}$U  &   0.65    & 0.67 & 0.64  & 0.90 &   0.85 & 0.92   \\
  $^{240}$Pu  &   0.65    & 0.67  & 0.57  & 0.85 &  0.84  & 0.76 \\
   $^{246}$Cm  &   0.70    & 0.66  & 0.58  & 0.85 &  0.83 & 0.85  \\
    $^{250}$Cf  &   0.67    & 0.65  & 0.64  & 0.84 & 0.83 & 0.78
\end{tabular}
\label{T2}
\end{table}

\narrowtext
\begin{table}
 \caption[A]{Size of the $Z$=114 and $N$=184 gaps
(in MeV) obtained for the nucleus $^{298}$114 within various
 models.}
\begin{tabular}{lcc}
 Model & 114 & 184  \\ \hline
HF+SLy7$^a$ & 1.55 & 2.92 \\
HF+SkP$^a$ &  0.63 & 2.36\\
WS$^a$ & 2.05  & 2.15\\
WS \protect\cite{[Nem72a]} & 2.5 & 2.14 \\
HF+SIII \protect\cite{[Bra74a],[Koh78]} & 1.93  & 1.6 \\
HF+SKa \protect\cite{[Koh78]} & 2.2  & 2.7  \\
BHF \protect\cite{[Koh71]} &  1.1 & 2.5 \\
BHF \protect\cite{[Cus76]} &  1.13 & 2.64 \\
ETF \protect\cite{[Ton80]} & 1.4  & 2.3 \\
RMF-NL1 \protect\cite{[Gam90]} & 1.47 & \\
RMF \protect\cite{[Boe93]} & 2.3 & 3.0 \\
DDRMF \protect\cite{[Boe93]} & 1.8 & 1.0
\end{tabular}
$^a$ This work.
\label{T114}
\end{table}

\narrowtext
\begin{table}
\caption[A]{Ground-state deformations
$\beta_2$ (upper number) and $\beta_4$ (lower number)
 for selected superheavy
nuclei predicted in the WS, FRDM, ETFSI, HF+SLy7, and HF+SkP models.}
\begin{tabular}{crrrrr}
Nucleus  & WS$^{\rm a}$~~\strut
 & FRDM$^{\rm b}$~~\strut & ETFSI$^{\rm c}$~~\strut
& HF+SLy7$^{\rm a}$~~\strut & HF+SkP$^{\rm a}$~~\strut
 \\ \hline
$^{264}108_{156}$ &    0.24 &    0.23 &    0.26 &    0.26 &    0.27 \\
                  & $-$0.04 & $-$0.04 & $-$0.05 & $-$0.05 & $-$0.04 \\[2mm]
$^{270}110_{160}$ &    0.23 &    0.22 &    0.25 &    0.25 &    0.25 \\
                  & $-$0.07 & $-$0.08 & $-$0.07 & $-$0.08 & $-$0.07 \\[2mm]
$^{278}112_{166}$ &    0.20 &    0.16 &    0.43 &    0.23 &    0.22 \\
                  & $-$0.10 & $-$0.06 &    0.07 & $-$0.10 & $-$0.09 \\[2mm]
$^{288}112_{176}$ &    0.09 & $-$0.06 &    0.43 &    0.11 &    0.12 \\
                  & $-$0.05 &    0    &    0.05 & $-$0.07 & $-$0.05 \\[2mm]
$^{298}114_{184}$ &    0    &    0    & $-$0.01 &    0    &    0    \\
                  &    0    &    0    &    0.04 & $-$0.01 &    0    \\[2mm]
$^{294}120_{174}$ &    0.08 &    0.08 &    0.43 &    0.09 &    0.10 \\
                  & $-$0.05 & $-$0.05 &    0.05 & $-$0.05 & $-$0.05 \\[2mm]
$^{294}122_{172}$ & $-$0.13 & $-$0.16 &    0.43 &    0.07 &    0.09 \\
                  &    0.01 &    0.03 &    0.05 & $-$0.03 & $-$0.04 \\[2mm]
$^{306}126_{180}$ & $-$0.24 &    0.42 &         &    0.01 &    0.01 \\
                  &    0.04 &    0.02 &         & $-$0.01 &    0    \\[2mm]
$^{310}126_{184}$ & $-$0.21 &    0    &         &    0    &    0    \\
                  &    0.02 &    0    &         & $-$0.01 &    0
\end{tabular}
\label{T3}
{$^{\rm a}$This work.}\\
{$^{\rm b}$Ref.~\protect\cite{[Mol95]}.}\\
{$^{\rm c}$Ref.~\protect\cite{[Abo95]}.}
\end{table}

\narrowtext
\begin{table}
\caption[A]{Calculated and experimental values
of $Q_\alpha$ for the $^{264}$Mt$\rightarrow$...$\rightarrow$$^{248}$Fm
$\alpha$-decay chain.
}
\begin{tabular}{ccccc}
Nucleus    & HF+SkP & HF+SLy7 & WS     & exp$^{\rm a}$ \\
\hline
$^{264}$Mt & 10.82  & 10.81  & 11.11  &     10.80   \\
$^{260}$Nh & 10.16  & 10.21  & 10.12  &     9.92   \\
$^{256}$Rf &  9.63  &  8.77  &  9.18  &     8.95   \\
$^{252}$No &  9.04  &  7.97  &  8.51  &     8.55
\end{tabular}
\label{CDATA}
{$^{\rm a}$Ref.~\protect\cite{[Aud95]}.}
\end{table}

\narrowtext
\begin{table}
\caption[A]{Theoretical estimates for the $\rmT_{\alpha}$ values
(in $\mu$s) for some nuclei}
\begin{tabular}{ccccc}
Nucleus    & HF+SkP & HF+SLy7 & WS     \\
\hline
$^{276}$112 & 4.3   & 0.5     &  4.7   \\
$^{280}$114 & 1.3   & 3.8     &  1.4   \\
$^{284}$116 & 2.4   & 49.0    &  0.5   \\
$^{288}$118 & 2.8   & 8.6     &  0.9
\end{tabular}
\label{TDATA}
\end{table}

\clearpage

\begin{figure}[ht]
\caption[A]{
Spherical  single proton levels in
$^{208}$Pb predicted in the
WS model as a function of the Coulomb potential strength $x_{\rm Coul}$
(for a definition see text).
The single-particle
levels are shifted by the energy $\Delta e$ given by Eq.~(\protect\ref{eshift})
with $V_{\rm Coul}(r$=0)=23.1597\,MeV.
}
\label{espPb}
\end{figure}

\begin{figure}[ht]
\caption[A]{
Same as Fig.~\protect\ref{espPb} but for single proton levels in
$^{208}126_{184}$. Here $V_{\rm Coul}(r$=0)=31.2890\,MeV.
}
\label{esp126A}
\end{figure}

\begin{figure}[ht]
\caption[esn114] {
Spherical neutron single particle levels in
$^{298}$114$_{184}$ predicted in the
HF+SkP, HF+SLy7, and WS models.
}
\label{esn114}
\end{figure}

\begin{figure}[ht]
\caption[esp114] {
Same as Fig.~\protect\ref{esn114} except for single proton levels
in  $^{298}$114.
}
\label{esp114}
\end{figure}

\begin{figure}[ht]
\caption[esn126] {
Same as Fig.~\protect\ref{esn114} except for single neutron levels
in  $^{310}$126.
}
\label{esn126}
\end{figure}

\begin{figure}[ht]
\caption [esp126]{
Same as Fig.~\protect\ref{esn114} except for single proton levels
in
$^{310}$126.
}
\label{esp126}
\end{figure}

\begin{figure}[ht]
\caption[A] {The macroscopic energy
$E_{\rm macro}$
(normalized to zero at
spherical shape) given by
the  Yukawa-plus-exponential  mass formula
as a function of $\beta_2$ for $^{264}$108, $^{298}$114, and $^{310}$126.
At every value of $\beta_2$, $E_{\rm macro}$ is minimized
with respect to deformations $\beta_4$ and $\beta_6$.
In addition to the instability to
prolate distortions seen in all three nuclei,
 there also develops an instability for $^{310}$126
with respect to oblate deformations.
}
\label{LD}
\end{figure}

\begin{figure}[ht]
\caption[A]
{Potential energy curve for
 $^{310}$126$_{184}$ obtained in the WS model
as a function of $\beta_2$. In order to emphasize shell effects, the
spherical part of the macroscopic energy, $E_{\rm macro}$, has been
substracted. At each value of $\beta_2$, the total energy $E_{\rm tot}$
(solid line)
has been minimized with respect to higher-multipole deformations
$\beta_4$ and  $\beta_6$. The macroscopic energy is also shown
(dashed line).
}
\label{landscape_WS}
\end{figure}

\begin{figure}[ht]
\caption[A]
{Potential energy curve for
 $^{288}$112$_{176}$ obtained with the Skyrme interaction SLy7
as a function of the quadrupole moment $Q_{20}$. The total energy
is normalized to zero at the spherical shape. Since the calculation
was not restricted to axial shapes, the triaxial deformation $\gamma$
varies as a function of  $Q_{20}$.
}
\label{landscape}
\end{figure}

\begin{figure}[ht]
\caption [b27]{
Contour map of the ground-state mass quadrupole deformation $\beta_2$
obtained with the Skyrme interactions SkP (top) and
SLy7 (bottom), plotted
as a function of $Z$ and $N$.
The deformation values at contour lines are indicated.
}
\label{b27}
\end{figure}

\begin{figure}[ht]
\caption [b47]{
Contour map of the ground-state mass hexadecapole
deformation $\beta_4$
obtained with the Skyrme interactions SkP (top) and
SLy7 (bottom), plotted
as a function of $Z$ and $N$.
The deformation values at contour lines are indicated.
}
\label{b47}
\end{figure}

\begin{figure}[ht]
\caption[A]{
Ground-state mass quadrupole
deformation $\beta_2$
obtained with the WS model plotted
as a function of $Z$ and $N$.
}
\label{shapes}
\end{figure}

\begin{figure}[ht]
\caption [qg112]{
Potential energy surface
of $^{288}$112
in the ($Q_{20}, Q_{22}$)-plane
calculated with the HF+SLy7 model.
The difference between contour lines is 0.5\,MeV. The local
minima and saddle points are indicated by dots and crossed dots, respectively.
}
\label{qg112}
\end{figure}

\begin{figure}[ht]
\caption [qg126]{
Same as in Fig.~\protect\ref{qg112}
except for the nucleus $^{310}$126 calculated with
the HF+SkP (top) and HF+SLy7 (bottom) models.
}
\label{qg126}
\end{figure}

\begin{figure}[ht]
\caption[esh] {
Shell energy, Eq.~(\protect\ref{Eshell}),  for SHE obtained with
HF+SkP (top), HF+SLy7 (middle), and WS (bottom) model.
Note different energy scale in the HF and WS calculations.
}
\label{esh}
\end{figure}

\begin{figure}[ht]
\caption[s2p] {Two proton separation energies
for even-even SHE obtained with
the  HF+SkP (top), HF+SLy7 (middle) and WS (bottom) models.
The difference between contour lines is 1\,MeV. The dashed lines
indicate the nuclei with  $S_{2p}<0$  unstable to the two-proton
emission. The positions of one-proton and  two-proton 
drip lines are indicated by thick solid and dotted lines, respectively.
}
\label{s2p}
\end{figure}

\begin{figure}[ht]
\caption[s2n] {Two neutron  separation energies
for even-even SHE obtained with
the  HF+SkP (top), HF+SLy7 (middle) and WS (bottom) models.
The difference between contour lines is 1\,MeV.
}
\label{s2n}
\end{figure}

\begin{figure}[ht]
\caption [qalpha]{
$Q_\alpha$ values for even-even SHE isotopes
obtained with the  HF+SkP (top),
HF+SLy7 (middle) and WS
(bottom) models. 
The difference between contour lines is 0.5\,MeV. Solid lines indicate
nuclei  with $Q_\alpha$$<$12.5\,MeV, i.e., those with
 alpha lifetimes greater than 1\,$\mu$s
 (see Sec.~\protect\ref{alpha} and  Fig.~\protect\ref{Tal}).
  }
\label{qalpha}
\end{figure}

\begin{figure}[ht]
\caption[betstab]{
The $\beta$-stability valley calculated with
the SkP (top) and the SLy7 (bottom) forces.
}
\label{betstab}

\end{figure}

\begin{figure}[ht]
\caption[Tal]{
Alpha decay half-lives, T$_\alpha$, for  even-even SHE isotopes
obtained with 
 the  HF+SkP,
HF+SLy7, and WS
models.
}
\label{Tal}
\end{figure}

\begin{figure}[ht]
\caption[tprot]{
One-proton emission half-lives
for even-even (solid line) and odd-even (dashed line) 
SHE
obtained in the WKB approximation
using the $Q_p$ values from
  the HF+SkP (left) and  HF+SLy7
(right) models and assuming $\ell$=0.
}
\label{tprot}
\end{figure}

\begin{figure}[ht]
\caption [qgo126]{
Potential energy curve 
for the nucleus $^{310}$126 obtained in the 
 HF+SkP (top) and HF+SLy7 (bottom) models
as a function 
of the mass quadrupole moment $Q_{20}$ 
with  (dashed line) and without (solid line) assumption
of the axial symmetry. The stronger shell effect at $Z$=126 
in SkP as compared to SLy7 lowers the gs configuration
in the former model. Consequently, the fission barrier
in SkP is higher.
}
\label{qgo126}
\end{figure}

\end{document}